\documentclass[12pt]{article}
\usepackage{latexsym}
\usepackage{amsbsy}

\newcommand{\dd}{\mbox{\rm d}}
\newcommand{\wg}{\wedge}
\newcommand{\gam}{\gamma}
\newcommand{\Gam}{\Gamma}

\newcommand{\ddg}{\ddagger}
\newcommand{\tl}{\tilde}

\newcommand{\DD}{\mbox{\rm D}}

\newcommand{\oo}{\over}
\newcommand{\p}{\partial}

\newcommand{\be}{\begin{equation}}
\newcommand{\bear}{\begin{eqnarray}}
\newcommand{\ear}{\end{eqnarray}}
\newcommand{\ee}{\end{equation}}
\newcommand{\lbl}{\label}
\newcommand{\ci}{\cite}

\newcommand{\vs}{\vspace}

\begin{document}

\

\baselineskip .6cm 

\vs{27mm}

\begin{center}

{\LARGE \bf
An Extra Structure of Spacetime: A Space of Points, Areas and
Volumes}\footnote{Talk presented at the XXIX Spanish Relativity
Meeting ERE 2006, 4th-8th September 2006, Palma de Mallorca, Spain}

\vs{3mm}

Matej Pav\v si\v c

Jo\v zef Stefan Institute, Jamova 39,
1000 Ljubljana, Slovenia

email: matej.pavsic@ijs.si

\vs{6mm}

{\bf Abstract}

\end{center}

\vs{2mm}

A theory in which points, lines, areas and volumes are on on the same
footing is investigated. All those geometric objects form a
16-dimensional manifold, called $C$-space, which generalizes spacetime.
In such higher dimensional space fundamental interactions can be unified
\` a la Kaluza-Klein. The ordinary, 4-dimensional, gravity and gauge fields
are incorporated in the  metric and spin connection, whilst the conserved
gauge charges are related to the isometries of curved $C$-space. It is
shown that a conserved generator of an isometry in $C$-space contains
a part with derivatives, which generalizes orbital angular momentum, and
a part with the generators of Clifford algebra, which generalizes spin.

\vs{8mm}

\section{Introduction}

In current approaches to quantum gravity the starting point is often
in assuming that at short distances there exists an underlying structure,
based, e.g., on strings and branes, or spin networks and spin foams\,
(see, e.g.,\ci{SpinNet}). 
It is then expected that the smooth spacetime manifold of classical
general relativity will emerge as a sufficiently good approximation at large
distances. However, it is feasible to assume that what we have, even at 
large distances, is in fact not just spacetime, but spacetime with certain
additional structure. The approach discussed in this contribution suggests
that the long distance approximation to a more fundamental structure 
is the space of extended events, i.e., points, lines, areas, 3-volumes,
and 4-volumes\,\ci{Pezzaglia}--\ci{PavsicKaluzaLong}.
All those objects can be elegantly represented \ci{Hestenes} by Clifford numbers
$x^M \gamma_M \equiv x^{\mu_1 ...\mu_r} \gamma_{\mu_1 ...\mu_r},~r=0,1,2,3,4$,
and therefore the corresponding space is called Clifford space ($C$-space).

It turns out that, since $C$-space is a higher dimensional space, it provides
a consistent description of quantized string theory\,\ci{PavsicSaasFee}.
The underlying 
spacetime can remain 4-dimensional, there is no need for a 26-dimensional or
a 10-dimensional spacetime. The extra degrees of freedom required
for consistency of string theory, described in terms of variables
$X^M (\tau, \sigma) \equiv X^{\mu_1 ...\mu_r} (\tau,\sigma)$, are due
to extra dimensions of $C$-space, and they need not be compactified;
they are due to the volume (area) evolution, and are thus physical.
But since a generic component $X^{\mu_1... \mu_r}$ denotes an 
oriented $r$-volume, associated with an $(r-1)$-brane
(i.e., a $p$-brane for $p=r-1$), we have that string itself (i.e., 1-brane)
is not enough for consistency. Higher branes are automatically present
in the description with functions $X^{\mu_1 ...\mu_r} (\tau,\sigma)$, 
although they are not described in full detail, but only up to the
knowledge of oriented $r$-volume. Because of the presence of two
parameters $\tau,~\sigma$, we keep on talking about evolving {\it strings},
not in 26 or 10-dimensional spacetime, but in 16-dimensional
Clifford space. In general, the number of parameters can be arbitrary,
but less then 16, and so we have a brane in $C$-space, i.e.,
a generalized brane\,\ci{PavsicBook}. Since with the points of a flat
$C$-space one can associate Clifford 
numbers (polyvectors), this automatically brings spinors (as members
of left or right ideals of Clifford algebra) into our description.
A polyvector $X^{\mu_1 ...\mu_r} \gamma_{\mu_1 ...\mu_r}$ can be rewritten
in terms of a basis spanning four independent left ideals, and thus contains
spinorial degrees of freedom.
This means that by describing our branes in terms of the $C$-space embedding
functions $X^M \equiv X^{\mu_1... \mu_r}$ we have already included spinorial 
degrees of freedom. We do not need to postulate them separately, as
in ordinary string and brane theories, where besides Grassmann even
( ``bosonic") variables $X^\mu$, there occur also Grassmann odd 
(``fermionic") or spinorial variables. In this formulation
we have a possible clue to the resolution of a big open problem,
namely, what exactly is string theory.

If we release the constraint of flat $C$-space, then we encounter 
a fascinating possibility, namely that curved $C$-space provides
a realization of Kaluza-Klein theory\,\ci{PavsicParis}--\ci{PavsicKaluzaLong},
and since all of its 16 dimensions are physically observable to us,
there is no need for compactification of the ``extra" dimensions.
The extra components of the $C$-space metric tensor are related to the
gauge fields that describe the fundamental interactions. When considering
the generalized Dirac equation in curved $C$-space it turns out that the
extra components of the $C$-space spin connection also manifest themselves
as gauge fields\,\ci{PavsicKaluza,PavsicKaluzaLong}. The conserved charges
consist of two contributions: one due to the ``orbital angular momentum",
and another one due to the ``spin"in
the ``internal" space. This is analogous to the situation that we know
from the ordinary Dirac theory: instead of spacetime we have $C$-space,
and instead of gamma ``matrices" we have now the full basis of Clifford
algebra. So instead of the ordinary angular momentum and spin, we
have the corresponding momenta in $C$-space, and in particular their
components in the ``internal" part of $C$-space, i.e., the part that
goes beyond spacetime.

\section{Extending spacetime to Clifford space}

In a manifold we do not have points only. We also have higher objects,
such as open and closed lines, surfaces, etc.\,. In a flat manifold
description of such objects can be done straightforwardly by employing
vectors of various grades: ordinary vectors, bivectors, trivectors, etc.,
in general $r$-vectors or multivectors, which are Clifford numbers.

Basis {\it vectors} (1-vectors) are generators of Clifford algebra
satisfying
\be
     \gam_\mu \cdot \gam_\nu \equiv \frac{1}{2} (\gam_\mu \gam_\nu + 
     \gam_\nu \gam_\mu) = g_{\mu \nu}
\lbl{2.1}
\ee
Higher grade vectors are given by the wedge product of vectors:
\bear
    \gam_\mu \wg \gam_\nu &\equiv& \frac{1}{2!}
    (\gam_\mu \gam_\nu - \gam_\nu \gam_\mu)
    \equiv {1\oo 2} [\gam_\mu,\gam_\nu]
  \lbl{2.2} \\
    \gam_\mu \wg \gam_\nu \wedge \gamma_\alpha &\equiv& \frac{1}{3!}
        [\gam_\mu,\gam_\nu,\gam_\alpha] \lbl{2.3} \\
        &\vdots&  \nonumber \\
   \gam_{\mu_1} \wg \gam_{\mu_2} \wg ... \, \wg \gam_{\mu_n} &\equiv& 
   \frac{1}{n!} [\gam_{\mu_1},\gam_{\mu_2},...,\gam_{\mu_n}]
 \lbl{2.3a}     
\ear

An oriented {\it line element} is
\be
     \dd x = \dd x^\mu \gam_\mu
\lbl{2.4}
\ee
where $\dd x^\mu,~\mu=0,1,2,3,...,$ are just the ordinary differentials of
coordinates (that should not be confused with the notation in the calculus
of differential forms). The $\dd x$ is a {\it vector}, expanded in terms
of {\it basis vectors} $\gam_\mu$, its components being $\dd x^\mu =
\dd x \cdot \gam^\mu$. Here $\gam^\mu$ are reciprocal basis vectors
satisfying $\gam^\mu \cdot \gam_\nu = {\delta^\mu}_\nu$.

Oriented {\it area element} is the wedge product of two line elements
$\dd x_1$ and $\dd x_2$:
\be
      \dd x_1 \wedge \dd x_2 = \dd x_1^\mu \dd x_2^\nu \, \gam_\mu \wg \gam_\nu
\lbl{2.5}
\ee
An arbitrary multivector element of degree $r$ is
\be
     \dd x_1 \wedge \dd x_2 \wg ...\wg \dd x_r
      = \dd x_1^{\mu_1} ... \dd x_r^{\mu_r} \, \gam_{\mu_1}\wg ... 
      \wg \gam_{\mu_r}
\lbl{2.6a}
\ee

At this stage we do not yet specify the dimensionality $n$ of the underlying
space(time); it can be arbitrary. The grade of a non vanishing $r$-vector
can be at most $r=n$.

If we are in flat spacetime manifold, we can straightforwardly integrate
the above infinitesimal $r$-vector line elements, and so we obtain
finite $r$-vectors $x^{\mu_1 ... \mu_r} \gam_{\mu_1 ...\mu_r}$ 
describing oriented r-dimensional areas ($r$-areas) associated
with {\it closed} $(r-1)$-dimensional surfaces\,\ci{PavsicArena},
or {\it open} $r$-dimensional
surfaces. We have thus introduced a manifold of $r$-areas, with
$x^{\mu_1 ... \mu_r}$ being their coordinates. The latter manifold 
is called {\it Clifford space}, or $C$-{\it space}.

This was just a manifold of abstract $r$-{\it areas} that generalizes
the ordinary manifold of abstract {\it points}. We can make contact
with physics by noting that in a spacetime manifold we have ``matter"
consisting of all sorts of physical objects that can be
point particles or extended objects, such as strings and branes of
various dimensionalities. A spacetime manifold can be considered as
a space of all possible positions that a test particle can have.

Analogously, an extended object can have many positions in a multidimensional
space of all possible object's configurations, the so called
{\it configuration space}. If extended objects are strings or branes,
then their configuration space is infinite dimensional. In refs.\,
\ci{PavsicBook} it was called ${\cal M}$-space. The latter space is
thus a space of all possible configurations that a (generalized) brane
can have. ${\cal M}$-space can have metric. Different choices of
${\cal M}$-space metric lead to different brane theories, with
different actions in the underlying spacetime. From the point of view
of spacetime we have different brane theories, but from the point of view
of ${\cal M}$-space there is one theory which allows for different background 
${\cal M}$-space metrics. The latter metric can become dynamical,
if we include into the description a corresponding kinetic
term\,\ci{PavsicBook}. Thus we arrive at a background independent
theory that generalizes general relativity to ${\cal M}$-space.

In order to avoid infinite dimensional description of branes, one can
introduce a quenched description\,\ci{AuriliaCastro,AuriliaQuenched}
working in a {\it finite} dimensional
subspace of ${\cal M}$-space. The finite dimensional space is the
space of oriented $r$-areas that we associate with closed $(r-1)$-branes,
or open $r$-branes. We have thus a many--to--one mapping from
the infinite dimensional objects, such as branes, to finite dimensional
$r$-areas. In other words, we have a many--to--one mapping from
infinite dimensional ${\cal M}$-space to finite dimensional
$C$-space\,\ci{Pezzaglia}--\ci{PavsicKaluzaLong}.

Clifford space is thus just a particular case of configuration space,
associated with an extended object. It takes into account not only
an object's center of mass position, but also its extension, which is
sampled by coordinates $x^{\mu_1 ...\mu_r}$.

In general, a ``point" in $C$-space can be described by coordinates
$x^{\mu_1 ...\mu_r}$. Two infinitesimally separated points, with
coordinates $x^{\mu_1 ...\mu_r}$ and $x^{\mu_1 ...\mu_r} +
\dd x^{\mu_1 ...\mu_r}$, define a {\it polyvector}
\be
    \dd X = \frac{1}{r!} \sum_{r=0}^n \dd x^{\mu_1 ...\mu_r}
    \gam_{\mu_1 ...\mu_r} \equiv \dd x^M \, \gam_M
\lbl{2.6}
\ee
which is a superposition of poyvectors of different grades, including
the scalars.

The line element in $C$-space is given by the quadratic
form\,\ci{Pezzaglia}--\ci{PavsicKaluzaLong}
\be
   \dd S^2 \equiv \dd X^\ddg * \dd X =
   \dd x^M \dd x^N G_{MN} \equiv \dd x^M \dd x_M
\lbl{2.7}
\ee
Here `*' denotes the scalar product of two polyvectors $A$ and $B$:
\be
     A*B = \langle A B \rangle_0
\lbl{2.8}
\ee
where $\langle ~~\rangle_0$ denotes the scalar part of the Clifford product
$AB$. Symbol `$\ddg$' denotes {\it reversion}, i.e., the operation
that reverses the order of vectors, e.g.,
\be
    (\gam_1 \gam_2 \gam_3)^\ddg = \gam_3 \gam_2 \gam_1
\lbl{2.9}
\ee
The metric is given by the scalar product of two basis Clifford numbers:
\be
   G_{MN} = \gam_M^\ddg * \gam_N
\lbl{2.10}
\ee
Reversion in the above definition is necessary for consistency reasons
\ci{PavsicSaasFee}.

If the signature of $n$-dimensional spacetime we start from is
pseudoeuclidean of the form $+ - - - ...\,$, then the signature of $C$-space
is $(p,p)$, with $p = 2^n/2$. This has important consequences for
string theory in particular, and for field theory in
general\,\ci{PavsicSaasFee,PavsicPseudoHarm}.

We can now envisage that physical objects are described
in terms of $x^M = (\omega, x^\mu, x^{\mu \nu},...)$. The
first straightforward possibility is to introduce a single
parameter $\tau$ and consider a mapping
\be
    \tau \rightarrow x^M = X^M (\tau)
\lbl{2.10a}
\ee
where $X^M (\tau)$ are 16 embedding functions that
describe a worldline in $C$-space. From the point of view of
$C$-space, $X^M (\tau)$ describe a wordlline of a ``point
particle": at every value of $\tau$ we have a {\it point} in
$C$-space. But from the perspective of the underlying
4-dimensional spacetime, $X^M (\tau)$ describe an extended
object, sampled by the center of mass coordinates $X^\mu (\tau)$
and the coordinates
$X^{\mu_1 \mu_2}(\tau),X^{\mu_1 \mu_2 \mu_3},
X^{\mu_1 \mu_2 \mu_3 \mu_4} (\tau)$.
They are a generalization of the center of mass coordinates in the sense
that they provide information about the object 2-vector, 3-vector, and
4-vector extension and orientation\footnote{A systematic and detailed
treatment is in refs. \ci{PavsicArena}.}.

The dynamics of such an object is determined by the action
\be
    I[X] = M \int \dd \tau \, ({\dot X}^{\dagger} *{\dot X})^{1\oo 2} =
   M \int \dd \tau ({\dot X}^M {\dot X}_M)^{1\oo 2}
\lbl{2.11}
\ee
The dynamical variables are given by the polyvector
\be
   X = X^M \gam_M = \Omega {\underline 1} + X^\mu \gam_\mu +
  X^{\mu_1 \mu_2} \gam_{\mu_1 \mu_2} + ... + X^{\mu_1 ...\mu_n} 
  \gam_{\mu_1 ... \mu_n}
\lbl{2.11a}
\ee
whilst
\be 
   {\dot X} = {\dot X}^M \gam_M = {\dot \Omega} {\underline 1} + 
   {\dot X}^\mu \gam_\mu +
  {\dot X}^{\mu_1 \mu_2} \gam_{\mu_1 \mu_2} + ... +
   {\dot X}^{\mu_1 ...\mu_n} \gam_{\mu_1 ... \mu_n}
\lbl{2.11b}
\ee
is the velocity polyvector, where ${\dot X}^M \equiv \dd X^M/\dd \tau$.

In the action (\ref{2.11}) we have a straightforward generalization of the
relativistic point particle in $M_4$:
\be
   I[X^\mu] = m \int \dd \tau ({\dot X}^\mu {\dot X}_\mu)^{1\oo 2}
  \; , \quad \mu = 0,1,2,3
\lbl{2.12}
\ee
 If a particle is extended, then (\ref{2.12}) provides only a very incomplete
description. A more complete description is given by the action
(\ref{2.11}), in which the $C$-space embedding functions $X^M (\tau)$
sample the objects extension.

\section{Strings and Clifford space}

Usual strings are described by the mapping $(\tau,\sigma) \rightarrow
x^\mu = X^\mu (\tau, \sigma)$, where the embedding functions
$X^\mu (\tau,\sigma)$ describe a 2-dimensional worldsheet swept
by a string. The action is given by the requirement that the area of the
worldsheet be ``minimal" (extremal). Such action is invariant under
reparametrizations of $(\tau,\sigma)$. There are several equivalent
forms of the action including the ``$\sigma$-model action" which, in
the conformal gauge, can be written as
\be
   I[X^\mu] = {\kappa\oo 2} \int \dd \tau \, \dd \sigma \, ({\dot X}^\mu
   {\dot X}_\mu - X'^\mu X'_\mu)
\lbl{3.1}
\ee
where ${\dot X}^\mu \equiv \dd X^\mu/\dd \tau$ and $X'^\mu \equiv
\dd X^\mu/\dd \sigma$. Here $\kappa$ is the string tension, usually
written as $\kappa = 1/(2 \pi \alpha')$.

If we generalize the action (\ref{3.1}) to $C$-space, we obtain
\be
     I[X] = {\kappa\oo 2} \int \dd \tau \, \dd \sigma \, ({\dot X}^M {\dot X}^N-
  X'^M X'_N)G_{MN}
\lbl{3.2}
\ee
Taking 4-dimensional spacetime, there are $D=2^4 = 16$ dimensions of $C$-space.
Its signature $(+++...- - - ...)$ has 8 plus and 8 minus signs.
This particular form of metric suggest to define vacuum according to
Jackiw et al.\,\ci{Jackiw1,Jackiw2}
(see also \ci{PavsicSaasFee,PavsicPseudoHarm}).
Then one finds that such generalized string theory is consistent.
There are no negative norm states, and the Virasoro algebra has no
central charges \,\ci{PavsicParis,PavsicSaasFee}).
My proposal is that, instead of adding extra dimensions to spacetime,
we can start from 4-dimensional spacetime $M_4$ with signature
$(+ - - -)$ and consider the Clifford space ${\cal C}_{M_4}$ ($C$-space)
whose dimensionality is 16 and signature $(8+,8-)$. {\it The necessary extra
dimensions for consistency of string theory are in $C$-space.} This
also automatically brings {\it spinors} into the game. It is an old
observation \ci{Riesz,Teitler}
 that spinors are the elements of left or right ideals of
Clifford algebras. In other words, spinors are particular sort of
polyvectors. Therefore, the string coordinate polyvectors contain spinors.
This is an alternative way of introducing spinors into the string theory.

\section{Curved Clifford Space}

As we can pass in the ordinary theory of relativity  from flat to curved
spacetime manifold, so we can pass from flat to curved
Clifford space $C$. This is a manifold such that at every point $X\in C$
its tangent space $T_X C$ is Clifford algebra, a vector space, whose
elements are polyvectors\,\ci{PavsicKaluzaLong}. Among them one can chose
those independent polyvectors which form a basis or frame. For variable $X$
we have a frame field.

It is convenient to distinguish two kinds of frame field:

~(i) {\it Coordinate frame field}
\be
   \gam_M \equiv \gam_{\mu_1 ... \mu_r}
\lbl{4.1}
\ee

(ii) {\it Orthonormal frame field}
\be
    \gam_A \equiv \gam_{a_1 ... a_r} = \gam_{a_1} \wg \gam_{a_2} \wg ...
      \wg \gam_{a_r}
\lbl{4.2}
\ee
While the basis elements $\gam_A$ are defined at every point point of
$C$-space as the wedge product of vectors, this is not the case for
$\gam_M$. The relation between $\gam_M$ and $\gam_A$ is given by
\be
    \gam_M = {e_M}^A \gam_A
\lbl{4.3}
\ee
where ${e_M}^A$ is the $C$-space vielbein. Eqs.\,(\ref{2.2})--(\ref{2.3a})
may hold at one point $X$, but not at different points of curved $C$ space.

At every point $X$ of $C$ a polyvector, an element of the tangent space
$T_X C$, can be expanded, e.g., im terms of the basis (\ref{4.1})
or (\ref{4.2}).
Since $X$ may run over the manifold $C$, we thus have a {\it polyvector field}.

At this point we encounter an important concept, namely, {\it the
derivative of a polyvector field}.

If acting  on a {\it scalar field}, the derivative is just the ordinary
{\it partial derivative}
\be
    \p_M \phi = {{\p \phi}\oo {\p x^M}}
\lbl{4.4}
\ee
If acting on a frame field, it defines {\it connection}, e.g., {\it
the connection for the coordinate frame field}
\be
     \p_M \gam_N = \Gam_{MN}^J \gam_J
\lbl{4.5}
\ee
or {\it the connection for the orthonormal frame field}
\be
    \p_M \gam_A = - {{\Omega_A}^B}_M \gam_B
\lbl{4.6}
\ee

If acting on a generic polyvector field, we have
\be
    \p_M (A^N \gam_N) = \p_M A^N \gam_N + A^N \p_M \gam_N =
    (\p_M A^N + \Gam_{MK}^N A^K) \gam_N \equiv \DD_M A^N \, \gam_N
\lbl{4.7}
\ee
The components $\DD_M A^N$ are {\it covariant derivative} of the tensor
analysis.

Contrary to the usual practice, we use in eqs.\,(\ref{4.4})--(\ref{4.7})
the same symbol $\p_M$. In ref.\,\ci{PavsicKaluzaLong} we argue in detail
why usage of different symbols for derivatives of different
geometric objects is unnecessary.

{\it Curvature} of $C$-space is defined, as usually, by the commutator
of derivatives acting on basis polyvecors:
\be
   [\p_M,\p_N] \gam_J = {R_{MNJ}}^K \gam_K
\lbl{4.8}
\ee
or
\be
   [\p_M,\p_N] \gam_A = {R_{MNA}}^B \gam_B
\lbl{4.9}
\ee
Introducing the reciprocal basis polyvectors $\gam^M$, $\gam^A$ satisfying
\be
    (\gam^M)^\ddg * \gam_N = {\delta^M}_N \; , \quad
    (\gam^A)^\ddg * \gam_B = {\delta^A}_B
\lbl{4.10}
\ee
we find from (\ref{4.8}) or (\ref{4.9}) the explicit expressions for
the components of curvature in the corresponding basis:
\be
    {R_{MNJ}}^K= \p_M \Gam_{NJ}^K - \p_N \Gam_{MJ}^K + 
    \Gam_{NJ}^L \Gam_{ML}^K -
   \Gam_{MJ}^L \Gam_{NL}^K
\lbl{4.11}
\ee
or
\be
    {R_{MNA}}^B= -(\p_M {{\Omega_A}^B}_N - \p_N {{\Omega_A}^B}_M
    + {{\Omega_A}^C}_N {{\Omega_C}^B}_M - {{\Omega_A}^C}_M {{\Omega_C}^B}_N)
\lbl{4.12}
\ee

A consequence of non vanishing curvature is that after the parallel
transport of a polyvector along a closed path we obtain a polyvector with
different orientation. In particular this means that, if initially
we have, e.g., a vector at a given point of $C$, then after a
round trip parallel transport to the same point we can obtain, e.g.,
a bivector, or in general any superposition of vectors, bivectors,
3-vectors, etc.\,.

Instead of 4-dimensional spacetime we have thus 16-dimensional $C$-space.
Since the latter space is higher dimensional, it can provide a realization
of Kaluza-Klein theory. Good features of $C$-space are:

\begin{description}

\item[\ \ (i)] There is no need for extra dimensions of spacetime.
Extra dimensions are in $C$-space.

\item[\ (ii)] There is no need to compactify the ``extra dimensions". The
extra dimensions of $C$-space, namely $\omega,~x^{\mu \nu},~x^{\mu \nu \rho},
~x^{\mu \nu \rho \sigma}$ sample the extended objects, therefore they
are physical dimensions.

\item[(iii)] The number of components $G_{\mu {\bar M}},~\mu=0,1,2,3,~
{\bar M} \neq \mu$, is 12, which is the same as the number of the
gauge fields in the standard model.

\end{description}

\section{Spinors as members of left ideals of Clifford algebra and
the generalized Dirac equation}

Let us consider a polyvector valued field $\Phi (X)$ on a curved $C$-space
manifold. At every point $X \in C$ a field $\Phi$ can be expanded,
e.g., in terms of the orthonormal basis according to
\be
    \Phi = \phi^A \gam_A\; , \qquad A=1,2,...,16
\lbl{5.1}
\ee
where $\phi^A$ are complex valued scalar components.

Alternatively, we can use another basis with elements $\xi_{\tl A} \equiv
\xi_{\alpha i} \in {\cal I}_i^L,~\alpha =1,2,3,4; i=1,2,3,4$, where
${\cal I}_i^L$ is the $i$-th left minimal ideal of Clifford
algebra\footnote{For a more detailed description see \ci{PavsicKaluzaLong}.}.
Expansion of a polyvector field then reads\,\ci{PavsicKaluza,PavsicKaluzaLong}
\be
    \Phi \equiv \Psi = \psi^{\tl A} \xi_{\tl A}
\lbl{5.2}
\ee
Such object is the sum of four independent 4-component spinors, each
in a different ideal ${\cal I}_i^L$.

By assumption a field $\Psi$ has to satisfy
\be
    \p \Psi \equiv \gam^M \p_M \Psi = 0
\lbl{5.3}
\ee
which generalizes the Dirac equation to $C$-space. The above equation is
{\it covariant}, because the derivative $\p_M$, if acting on a generalized
spinorial basis elements $\xi_{\tl A}$, gives the generalized spin
connection:
\be
    \p_M \xi_{\tilde A} = {{\Gam_M}^{\tilde B}}_{\tl A} \xi_{\tl B}
\lbl{5.4}
\ee
Using the latter relation, we can write eq.\,(\ref{5.3}) in the form
\be
    \gam^M \p_M (\psi^{\tl A} \xi_{\tl A}) = \gam^M (\p_M \psi^{\tl A}
  + {{\Gam_M}^{\tl A}}_{\tl B} \psi^{\tl B}) \xi_{\tl A} 
    \equiv \gam^M (\DD_M \psi^{\tl A})\, \xi_{\tl A} = 0
\lbl{5.5}
\ee

An action which leads to eq.\,(\ref{5.3}) is\,\ci{PavsicKaluzaLong}:
\be
    I[\Psi, \Psi^\ddg] = \int \dd^{2^n} x \, \sqrt{|G|} \, 
    i \Psi^\ddg \p \Psi = \int \dd^{2^n} x\, \sqrt{|G|} \, i
   {\psi^*}^{\tl B} \xi_{\tl B}^\ddg \gam^M \xi_{\tl A} \DD_M \psi^{\tl A}
\lbl{5.6}
\ee
where $\dd^{2^n} x \, \sqrt{|G|}$ is the invariant volume element of the
$2^n$-dimensional $C$-space, $G\equiv {\rm det}\, G_{MN}$. We take $n=4$.

A generic transformation in the tangent $C$-space $T_X C$ which maps
a polyvector $\Psi$ (i.e., a generalized spinor)
into another polyvector $\Psi'$ is given by
\be
   \Psi' = R \Psi S
\lbl{5.7}
\ee
where  $R = {\rm e}^{{1\oo 4} \Sigma_{AB} \alpha^{AB}}$ and 
$S= {\rm e}^{{1\oo 4} \Sigma_{AB} \beta^{AB}}$,
with
$\alpha^{AB}$ and $\beta^{AB}$ being parameters of the left and right
transformations, respectively. The generators $\Sigma_{AB}$ are
defined as
\be
      \Sigma_{AB} = - \Sigma_{BA}
      = \left\{  \begin{array}{ll}
      ~~\gam_A \gam_B\; , & {\rm if} ~A < B \\
      - \gam_A \gam_B\; , & {\rm if} ~A > B \\
      ~~~0 \; , & {\rm if} ~A = B
      \end{array}
      \right.
\lbl{5.8}
\ee
Transformation (\ref{5.7}) can be written in the form of a $16\times 16$
matrix:
\be
    \psi'^{\tl A} = {U^{\tl A}}_{\tl B} \psi^{\tl B}
\quad {\rm or} \quad
    \psi' = {\bf U} \psi
~~, \qquad {\bf U} = {\bf R} \otimes {\bf S}^{\rm T}
\lbl{5.9}
\ee
where ${\bf R}$ and ${\bf S}$ are
$4 \times 4$ matrices representing the Clifford numbers $R$ and $S$.
We see
that  ${\bf U}$ is the direct product of ${\bf R}$ and
the transpose  ${\bf S}^{\rm T}$
of ${\bf S}$.

Under the transformation (\ref{5.9}) the generalized spin connection
transforms as
\be
    {\Gamma_{M {\tl A}}}^{\tl B} = {U_{\tl D}}^{\tl B}
     {U^{\tl C}}_{\tl A} {\Gamma'_{M {\tl C}}}^{\tl D} +
      \p_M {U^{\tl D}}_{\tl A}\, {U_{\tl D}}^{\tl B}
\; ,
      \quad {\rm i.e.}, \quad
      {\bf \Gamma}_M = {\bf U}\, {\bf \Gamma'_M}\, {\bf U}^{-1} + 
    {\bf U}\, \p_M \, {\bf U}^{-1}
\lbl{5.10}
\ee
while the covariant derivative transforms as
\be
 {\DD}'_{M} \psi'^{\tl A} = {U^{\tl A}}_{\tl B} \, \DD_M \psi^{\tl B}
\;, \quad {\rm i.e.}, \quad
\DD'_M \psi' = {\bf U}\, \DD_M \psi
\lbl{5.11}
\ee 
where $\DD'_{M}\psi'^{\tl A} = \p'_{M} \psi'^{\tl A} +
      {{\Gam'_{M}}^{\tl A}}_{\tl B} \psi'^{\tl B}$ and
      $\DD_{M}\psi^{\tl A} = \p_M \psi^{\tl A} +
      {{\Gam_{M}}^{\tl A}}_{\tl B} \psi^{\tl B}$.
We see that ${\bf \Gam}_M$ transforms as a non abelian gauge field.
 
The generally covariant equation in 16-dimensional $C$-space contains
the coupling of spinor fields $\psi^{\tl A}$ to non abelian gauge
fields ${{\Gam'_{M}}^{\tl A}}_{\tl B}$ which altogether form components
of connection in the generalized spinor basis.

\section{Conserved charges and isometries}

In curved space in general there are no conserved quantities, unless
there exist isometries which are described by Killing vector fields.
Suppose we have a curved Clifford space which admits $K$ Clifford numbers
$k^\alpha = k_M^\alpha\, \gam^M\,$,  ~$\alpha = 1,2,...,K;~ M=1,2,...,16$,
where the components satisfy the condition for isometry, namely
\be
     \DD_N k_M^\alpha + \DD_M k_N^\alpha = 0
\lbl{6.1}
\ee
the covariant derivative being defined in eq.\,(\ref{4.7}). 
We assume that such $C$-space with isometries is not given ad hoc, but
is a solution to the generalized Einstein equations that arise from the action
which contains a ``matter" term, such as (\ref{5.6}), and a field term 
defined by means of the $C$-space curvature
(see\,\ci{PavsicKaluzaLong}).

Taking a coordinate system in which $k^{\alpha \mu}=0,~k^{\alpha {\bar M}}
\neq 0,~\mu=0,1,2,3,~{\bar M} \neq \mu$, the metric and
the
vielbein can be written as\footnote{This is a $C$-space analog of the
Kaluza-Klein splitting usually performed in the literature. See, e.g.,
\ci{Luciani,Witten}.}
\be
   G_{MN} = \pmatrix{
            G_{\mu \nu} & G_{\mu {\bar M}}\cr
            G_{{\bar M} \nu} & G_{{\bar M}{\bar N}}\cr
            } \; , \quad 
    {e^A}_M = \pmatrix{
            {e^a}_\mu & {e^a}_{\bar M} \cr
            {e^{\bar A}}_\mu & {e^{\bar A}}_{\bar N} \cr
            } 
\lbl{6.2}
\ee
Here
\be
    {e^a}_{\bar M} = 0
\lbl{6.3}
\ee
whilst the components ${e^{\bar A}}_\mu$ can be written in terms of Killing
vectors and gauge fields $W_\mu^\alpha (x^\mu)$:
\be
    {e^{\bar A}}_\mu = {e^{\bar A}}_{\bar M} \, 
    k^{\alpha {\bar M}} W_\mu^\alpha \; , \quad
    \quad \p_{\bar M} W_\mu^\alpha = 0
\lbl{6.4}
\ee
If we set the $C$-space torsion to zero and calculate
the connection $\Omega_{ABM}$ by
using eqs.\,(\ref{6.2})--(\ref{6.4}), we obtain an analogous result
as given, e.g., in ref.\,\ci{Luciani}:
\be
    \Omega_{{\bar M}{\bar N} \, \mu} = 
    {1\oo 2} k_{[{\bar M},{\bar N}]}^\alpha W_\mu^\alpha
\lbl{6.5}
\ee
where $ k_{[{\bar M},{\bar N}]}^\alpha \equiv
\p_{\bar N} k_{\bar M}^\alpha -\p_{\bar M} k_{\bar N}^\alpha$

Let us now rewrite the $C$-space Dirac equation by using
eqs.\,(\ref{6.1})--(\ref{6.5}). Omitting the terms due to the
$C$-space torsion, we obtain
\be
    \left [ {{\bf \gam}^{(4)}}^\mu 
    \left ( \p_\mu - \Omega_{ab \,\mu} {1\oo 8} 
    [{\bf \gam}^a,{\bf \gam}^b] - q^\alpha \, W_\mu^\alpha
     + ...\right ) + 
    {\bf \gam}^{\bar M} \p_{\bar M} + ... \right ] \psi
= 0
\lbl{6.6}
\ee
where ${{\bf \gam}^{(4)}}^\mu = {e_a}^\mu {\bf \gam}^a$ are 4-dimensional
coordinate basis vectors, and
\be
     q^\alpha = k^{\alpha \, {\bar M}} \, \p_{\bar M} + {1\oo 8}\, 
      k_{[{\bar M},{\bar N}]}^\alpha \,{e_{\bar A}}^{\bar M}
      {e_{\bar B}}^{\bar N}\, \Sigma^{{\bar A}{\bar B}}
\lbl{6.7}
\ee
are the charges, conserved due to the presence of isometries
$k^{\alpha \, {\bar M}}$. They are the sum of the coordinate
part and the contribution of the spin angular momentum in the ``internal"
space, spanned by $\gam^{\bar M}$. The coordinate part is the projection
of the linear momentum onto the Killing (poly)vectors, and can in particular be
just the orbital angular momentum of the ``internal" part of $C$-space.
The first term that contributes to the charge $q^\alpha$ comes from
the vielbein  according to eq.\,(\ref{6.4}), whilst the second term comes
from the connection according to eq.\,(\ref{6.5}). Both terms couple
to the same
4-dimensional gauge fields $W_\mu^\alpha$, where the index
$\alpha$ denotes which gauge field (which Killing polyvector), and should not
be confused with the spinorial index, used in Dirac matrices.

In eq.\,(\ref{6.6}) we explicitly displayed only the the most relevant
terms which contain the the ordinary
vierbein ${e^a}_\mu$ and spin connection $\Omega_{ab \,\mu}$
(describing gravity and torsion), and also Yang-Mills gauge fields
$W_\mu^\alpha$ which, as shown in eqs.\,(\ref{6.4}),(\ref{6.5}), occur
in the $C$-space vielbein and in the $C$-space ``spin" connection.
We omitted the terms arising form the $C$-space torsion

\section{Discussion}

Clifford space provides a promising approach to the unification of
fundamental interactions. At first sight one might think that the
signature $(p,p)$ brings ghosts into the description. This is not the
case, if we adopt the Jackiw et al. definition of
vacuum\,\ci{Jackiw1,Jackiw2,PavsicPseudoHarm,
PavsicSaasFee}. Then such pseudoeuclidean signature turns out to be
welcome for string theory\,\ci{PavsicSaasFee} and for the resolution of
the cosmological constant problem\,\ci{PavsicPseudoHarm}.

Another possible obstacle could be seen in the Coleman-Mandula theorem
\ci{ColemanMandula} which forbids nontrivial mixing of spacetime
and internal symmetries. But all such theorems are based on certain,
often tacit, assumptions. One such tacit assumption (technically expressed
in terms of certain properties of $S$-matrix) in the derivation of the
Coleman-Mandula theorem is that probability conservation refers to
spacetime, and not to the ``internal" space. In other words, the internal
space was not taken on equal footing as spacetime at the very beginning.
However, it was shown by Pelc and Horwitz\,\ci{HorwitzMandula} that
Coleman-Mandula theorem can be extended to a higher dimensional space.
In general, in Kaluza-Klein approach 
conserved generators associated with gauge charges come from the
extra dimensions. Extra dimensions have in principle the same role
as the ordinary four dimensions, and there are the transformations
that transform one into the other.  Thus internal symmetries, associated with
extra dimensions, can in general nontrivially  mix with four spacetime
dimensions. Analogously holds for Clifford space. In particular,
however, a curved higher dimensional space can admit isometries, and can
then be written as the direct product of 4-dimensional spacetime and
the internal space. We have shown that the conserved
charges, which are due to the isometries of curved Clifford space, have two
contributions: one from the  ``orbital'' angular momentum in the
``internal" part of $C$-space, and the other from the ``internal" spin.
This could have some important physical consequences that need to be
explored further.

\end{document}